The Inhibition of B16-F10 Mouse Melanoma Cell Proliferation in Increased Concentration Levels of Theophylline

Faiza Hasib Khan

Megan Robles

Gettysburg College Biology Department

Dr. Hiraizumi

# ABSTRACT

Cancer is a complex sequence of disease conditions progressing gradually with generalized loss of growth control. It continues to be one of the biggest global health problems, and its etiology has given rise to a huge array of treatment outside of conventional chemotherapy. Melanoma is one of the deadliest forms of skin cancer originating from melanocytes. It is characterized by the overproduction of melanin by the increased in cell proliferation. Melanogenesis, the production of melanin is by melanocyte-stimulating hormone (MSH) which stimulates cyclic AMP (cAMP) production to increase melanocyte production. Through the use of methylxanthines, theophylline proliferation rate can be decreased by increasing cell differentiation. One of the basic principles of cell biology is the selectivity of differentiation and proliferation, where cells usually grow or differentiate but not both. This study aimed to collect baseline data of untreated B16-F10 melanoma cells to determine morphology and doubling time of untreated cells. This data was further compared to results from varied concentration treatments of theophylline. It was hypothesized that increased levels in cyclic adenosine monophosphate (cAMP) could induce differentiation of melanocytes to terminate proliferation. To test this hypothesis, we collected a series of images of B16- F10 derived from mouse melanoma cells and seeded cells into a 12-well plate to calculate cell concentration for a five-day period and doubling time for untreated cells. Theophylline levels were varied to stimulate the production of cAMP and determine its effect on melanocyte proliferation and differentiation. The observed results showed that intracellular stabilization of cAMP, via phosphodiesterase inhibition by methylxanthines like theophylline, increases cell differentiation on melanoma cells and suppresses their growth.

# INTRODUCTION

Cancer is a complex sequence of disease conditions progressing gradually with generalized loss of growth control (Debela et al., 2020). It continues to be one of the biggest global health problems responsible for one in six deaths worldwide (Debela et al., 2021). In 2020, 19.3 million new cancer cases and about 10 million cancer deaths globally were estimated (Debela et al., 2020). Its etiology has given rise to a huge array of treatments outside of conventional chemotherapy.

Conventional treatment approaches such as surgery, chemotherapy and radiotherapy are commonly use while advances in new cancer therapy treatments are being develop (Debela et al., 2020) Chemotherapy acts on dividing cells by inducing DNA damage and strand breakage, interfering with DNA damage repair and microtubule function. (Stone and DeAngelis, 2016). It is considered one of the most effective and widely used modalities in treating cancer as used alone or in combination with radiotherapy (Debela et al., 2020). Radiation therapy, similarly, targets dividing cells and can cause damage to neural structures directly or indirectly by causing vascular damage and fibrosis of neural structures (Stone and DeAngelis, 2016).

Many pathways involved in cancer therapy progression and how they can be targeted has improved drastically in response to a combination of strategies that involves multiple target therapies or traditional chemotherapeutics, such as taxanes and platinum compounds (Debela et al., 2020). New approaches, such as drugs, biological molecules, and immune-mediated therapies have been used as a method of treatment despite the inefficiency of conventional therapies to reach the point where it resists the mortality rate, while extending the duration of survival for metastatic cancer (Debela et al., 2020).

Drug resistance and its delivery systems are the most problem in cancer cure and decreasing sign and symptoms (Debela et al., 2020). Currently, advances are being made in many approved treatment approaches and drugs (Debela et al., 2020). One promising approach is differentiation therapy, which aims to drive cancer cells into a mature, specialized state rather than simply eliminating them through cytotoxicity (Breitman et al.,1980). Instead of only killing the cancer cells, differentiation therapy seeks to induce them to return to a differentiated, non-dividing state. (Breitman et al.,1980). This is an encouraging prospect by causing cells to take on mature, specialized functions. This method is intended to avoid the aggressive growth characteristic of cancer and possibly limit malignancy without the harsh side effects of cytotoxic chemicals. (Breitman et al.,1980).

Stem cells are undifferentiated cells present in bone marrow that have the ability to differentiate into any type of body cell (Debele et al., 2020) Stem cell therapeutic strategy is considered to be a safe and effective treatment option for cancer. The application of stem cells in the experimental clinical trial, such that the regeneration of other damaged tissue is being explored (Debele et, al., 2020).

Targeted therapy is another kind of cancer therapy targets a particular region, such as intracellular organelles or tumor vasculature, without affecting adjacent tissues (Kaur et al., 2023) It works by interfering with the growth of molecules to block cancer growth and spreading (Debele et, al., 2020). It differs from standard chemotherapy by attacking cancer cells with less damage to normal cells, focusing on characteristics that set cancer cells apart (Debele et, al., 2020).

In this study, we focused on an important public health concern, the type of cancer. Melanoma is one of the deadliest forms of skin cancer (Guy et al.,2012). It is caused by the malignant transformation of melanocytes, which are cells in cutaneous tissue that produce melanin

(Lauters et al, 2024). The incidence of melanoma has been drastically increasing. In 2008 there were 59,695 newly diagnosed cases and 8,623 deaths from melanoma in the U.S. (Guy et al.,2012). The average lifetime risk of developing melanoma in the U.S. has increased significantly from 1 in 1500 in 1935 to 1 in 30 in 2009 (Guy et al.,2012).

Melanocytes are pigment-producing cells found in hair follicles and the basal layer of the epidermis. They do this by synthesizing the protective pigment melanin and delivering it to the keratinocytes, the majority of the epidermis' cells. Production of melanin starts when the enzyme tyrosinase converts the amino acid tyrosine into melanin. The pigment is subsequently loaded into specialized organelles called melanosomes. These melanosomes defend skin cells against UV radiation by moving actively on dendritic extensions and thereafter being released and endocytosed into keratinocytes. Intracellular signaling molecule cyclic AMP (cAMP) is intimately related to cellular differentiation of melanocytes.

Intracellular cAMP levels are elevated due to the activity of melanocyte-stimulating hormone Melanoma, a highly aggressive cancer arising from melanocytes, provides a compelling model for investigating differentiation therapies (Gilchrest et al., 1999; Miller and Mihm, 2006). Activation of a G-protein coupled receptor (GPCR) by melanocyte-stimulating hormone (MSH) leads to the activation of adenylyl cyclase, raising intracellular cAMP levels (Alberts et al., 2023). By activating melanin production and the growth of dendrites and by simultaneously inhibiting cell growth, elevated cAMP levels thus induce melanocyte differentiation. One of the basic principles of cell biology is the selectivity of differentiation and proliferation, where cells usually grow or differentiate but not both. This assumption is the foundation of the reasoning of differentiation therapy in oncology. Pharmacological agents such as methylxanthines, including theophylline, can elevate intracellular cAMP levels by inhibiting phosphodiesterase, the enzyme responsible for cAMP degradation (Tada et al., 2002). They can induce cell differentiation and inhibit cancer cell growth by stabilizing high levels of cAMP and offer a likely treatment strategy for melanoma and other cancers.

Three principal aims were considered in this experiment: first, to describe the morphology and proliferation of untreated B16-F10 melanoma cells in culture; second, to establish if treatment with theophylline changes the proliferation rate of the cells; and third, to establish if an increase in cAMP levels using theophylline causes morphological differentiation. In this study we used MTT assay as an indirect indicator of cell proliferation and phase contrast microscopy to determine morphological change in different conditions with varied levels of theophylline, a methylxanthine treatment to analyze its effects on cell differentiation. We could evaluate the promise of cAMP elevation as a model for differentiation therapy in melanoma cells by integrating these methods.

## MATERIALS AND METHODS

<u>Photographing untreated B16-F10 cells</u>

Cells seeded into T-flasks with sufficient time to reach sufficient confluence to establish the baseline morphology of untreated melanoma cells were provided by the lab instructor. Phase contrast microscopy was conducted on an EVOS 5000 inverted microscope, and representative images were taken at 400x magnification to view important morphological features such as dendrites, lamellipodia, mitotic figures, nucleoli, and Golgi apparatus. Mouse Melanoma cells CRL-6475) were provided by the ATCC and subsequently cultured in Dulbecco's Modified Eagle Medium (DMEM; Gibco) with 10% fetal calf serum (FCS; Thermo Fisher Scientific) and 1% penicillin-streptomycin (Gibco).

Seeding Cells intro 12-Well Plates

100µl stock cells B16-F10 derived from a mouse melanoma with a cell concentration of approximately $1 \times 10^6$ cell/ml provided by the lab instructor were added to a 15 ml conical vial containing 10ml of sterile culture medium (DMEM + 10% FCS + 1% pen/strep). 2 ml of the cell's suspension was added into 6 wells of a sterile 12-well plate. Cell was incubated in an atmosphere at 5% $CO_2$ at 37˚C.

Counting Cells with Hemocytometer

Cell counts were taken daily for a period of five days using sterile and standard hemocytometer technique. The cell concentration per well was calculated by using the mean cell concentration of $5.23 \times 10^6$ cell/ml obtained from cell counting on day cero of the experiment. On the following days (1-5), DMEM was removed from the well and cells were washed with 1ml of phosphate-buffered saline (PBS). PBS was retrieved and 200µl of Accutase solution was added. Cells were placed on the incubator for 35 minutes. After the 35-minute incubation period, the cell susception was triturated and placed on a sterile microfuge tube containing 200µl of 0.4% Trypan blue in PBS. Viable cells were counted at intervals of 22 - 28 hours over the course of 134 hours.

Theophylline treatment

100µl of stock B16-F10 cells were added to a 15ml conical vial with various amounts of medium and theophylline (obtained from Sigma-Aldrich). In the control medium 100µl of medium were removed from the 15ml vial containing 10 ml of medium, and 100µl of stock B16-F10 cells were added. In the low theophylline concentration treatment, 200µl of medium were removed from the 15ml vial containing 10 ml of medium, 100µl of stock B16-F10 cells, and 200µl of theophylline were added for a final concentration of 1mM. In the high theophylline concentration condition,

1000µl of medium were removed from the 15ml vial containing 10 ml of medium, 100µl of stock B16-F10 cells, and 1000µl of theophylline were added for a final concentration of 5mM. 2 ml of the cell suspension for each treatment were added to a 12-well plate, each treatment had a total of four replicates. After 48 hours two representative photos were taken for each treatment using the EVOS 5000 microscope with a total magnification 200x. MTT assay was performed on day three and plate was stored on the refrigerator for five days. The absorbance was measured on day nine with the assistance of the lab instructor using a spectrophotometer reader (BioTek Epoch or similar).

Dimethylthiazol tertrazolium (MTT) Assay

Medium was retrieved from all wells and replaced with 900µl of Minimal Essential Medium without serum or phenol red (MEMminus). 100µl of a 10x stock solution of MTT (5mg/ml) was added into each well, and the plate was incubated for a 3-hour period in an atmosphere at 5% $CO_2$ at 37˚C. MEMmimus was removed from each well and formazan was dissolved by adding 1ml of isopropanol. Plate was stored in the refrigerator for six days before measuring absorbance at 570 nm at 25˚C.

Data Analysis

*Doubling Time*

To determine the doubling time of untreated B16-F10 mouse melanoma, a semilog plot of cell number per well against time (hours) of the data from the growth curve (Figure 2). Data points on the lag and stationary phase of the curve were omitted for a linear regression with the highest $R^2$ value ($R^2 = 1$). Doubling time was calculated by diving log 2 by the slope of the linear regression equation, y = 0.4459x – 34.442.

*Absorbance*

True absorbance values were calculated by subtracting 0.048 from each well absorbance value and multiplying it by four. The mean, standard deviation, and standard error of the mean were calculated. A one-way analysis of variance (ANOVA) on the data was conducted to determine if there was a statistical difference between experimental treatments.

**RESULTS**

Photographs of untreated B16-F10 cells derived from a mouse melanoma (Figure 1) and doubling time of untreated cells (Figure 3) data was collected in weeks one and two of the experiment. The absorbance data from the MTT Assay obtained during week three of the experiment was further analyzed by performing a statistical analysis in form of a one-way ANOVA test ($p = 0.544376$). No statistically significant difference was found between the different conditions. The mean absorbance values for the different concentration treatments, 0mM, 1mM, and 5mM were 0.824, 0.52 and 0.231, respectively.

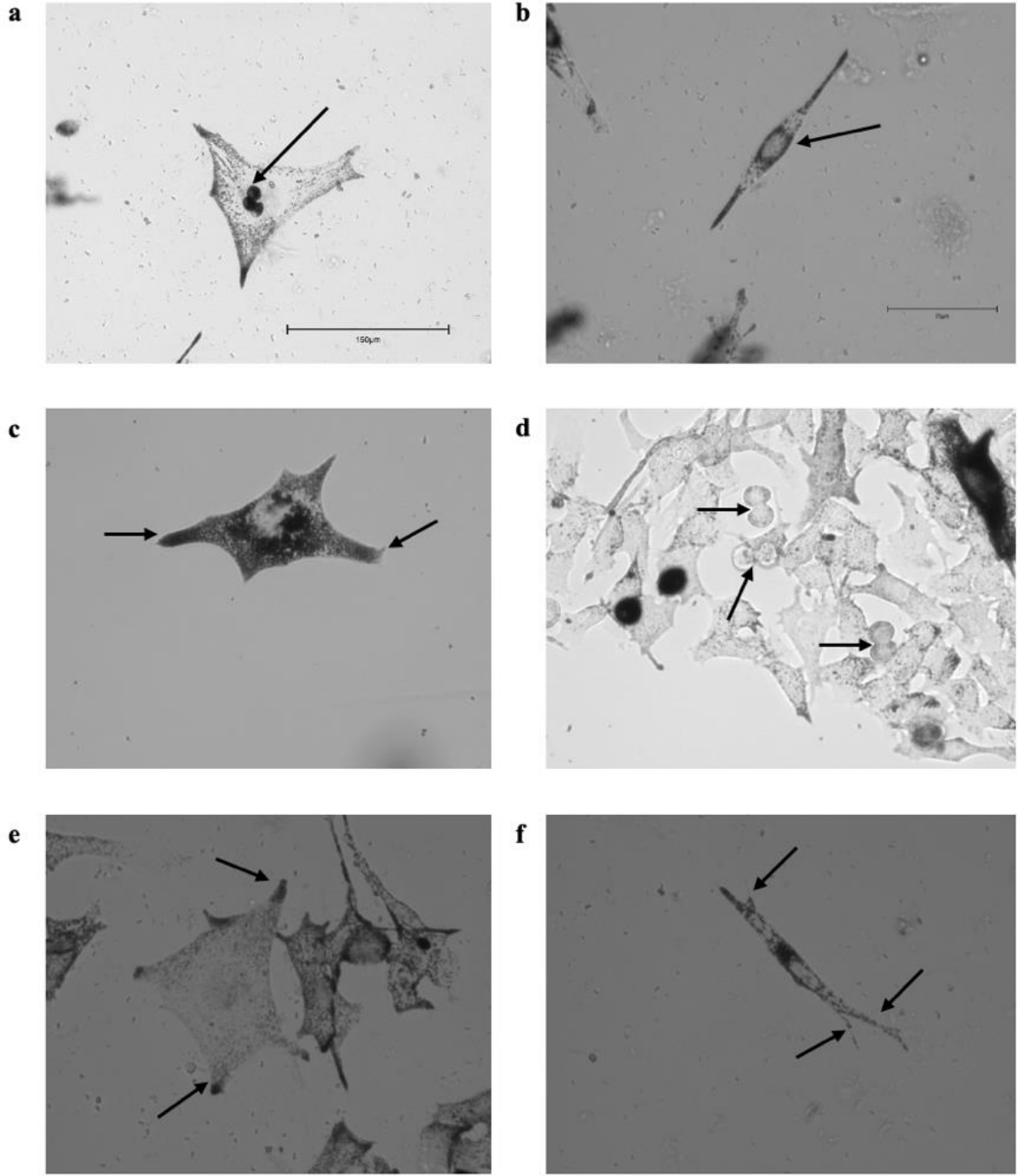

**Figure 1.** Photographs of unique, untreaded B16-F10 cells derived from mouse melanoma. Photographs were taken using the EVOS 500 microscope at a total magnification of 400x for a, b, c, e and f, a total magnification of 200x was used for d. Image a show a cell with different numbers of nucleoli. Image b shows a cell with a distinct Golgi apparatus next to the nucleus. Image c and e show crawling cells with lamellipodia. Image d shows mitotic cells in telophase of mitosis. Image f shows a cell with multiple dendrites (refer to the appendix for timelapse of miotic cell division). Arrows emphasize the characteristics mentioned in each photograph.

To determine the doubling time of a normal cell culture, a growth curve of untreated B16-F10 mouse melanoma cells was created (Figure 2) using data collected for a period of 134 hours with cell counts taken every 22 to 28 hours. The mean cell concentration was calculated to be 5.23 x $10^6$ cell/ml. A semilog plot of the of the data collected was created (Figure 3) to obtain the doubling time of a normal culture cells.

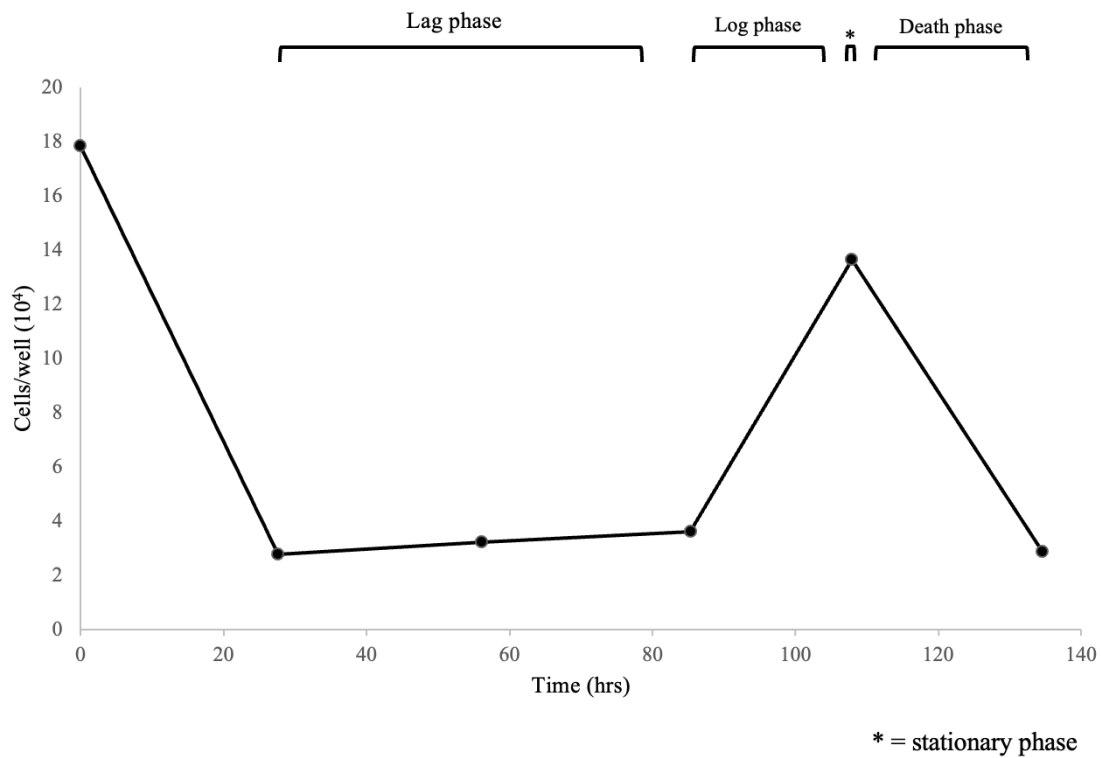

**Figure 2**. Growth curve of B16-F10 cells over the course of 134-hour period with cells counts taken about every 22 - 28 hours using a hemocytometer. Graph illustrates a lag phase between the 27 to 56 hours period, followed by a period of a 22-hour of exponential phase. The stationary phase was present for a short period of time, followed by death phase. Original mean cell concentration was determined to be 5.23 x $10^6$ cell/ml.

To identify the timeline with the most exponential cell growth to determine the doubling time of untreated cells. The data points of the lag and stationary phase were omitted (Figure 2) to obtain the highest $R^2$ value for a linear regression.

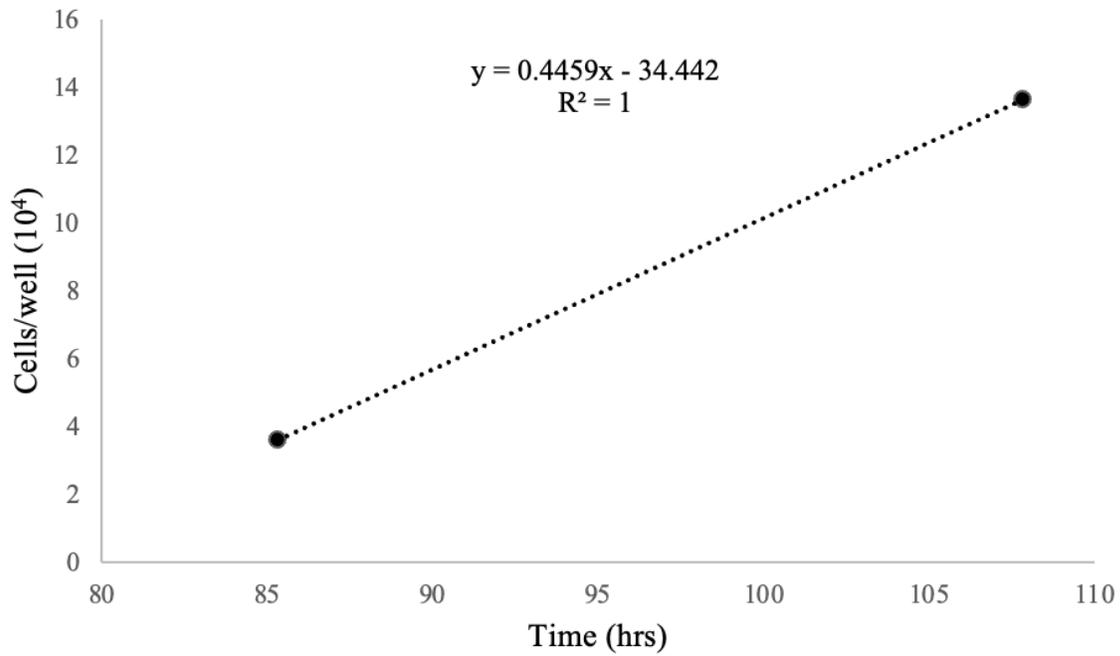

**Figure 3.** Semilog graph with the highest $R^2$ value for a linear regression from the logarithmic phase of the growth curve (figure 2)/ Linear regression equation y = 0.4459x – 34.442; $R^2 = 1$ was used to calculate doubling time. The doubling time was calculated to be 0.67 hours.

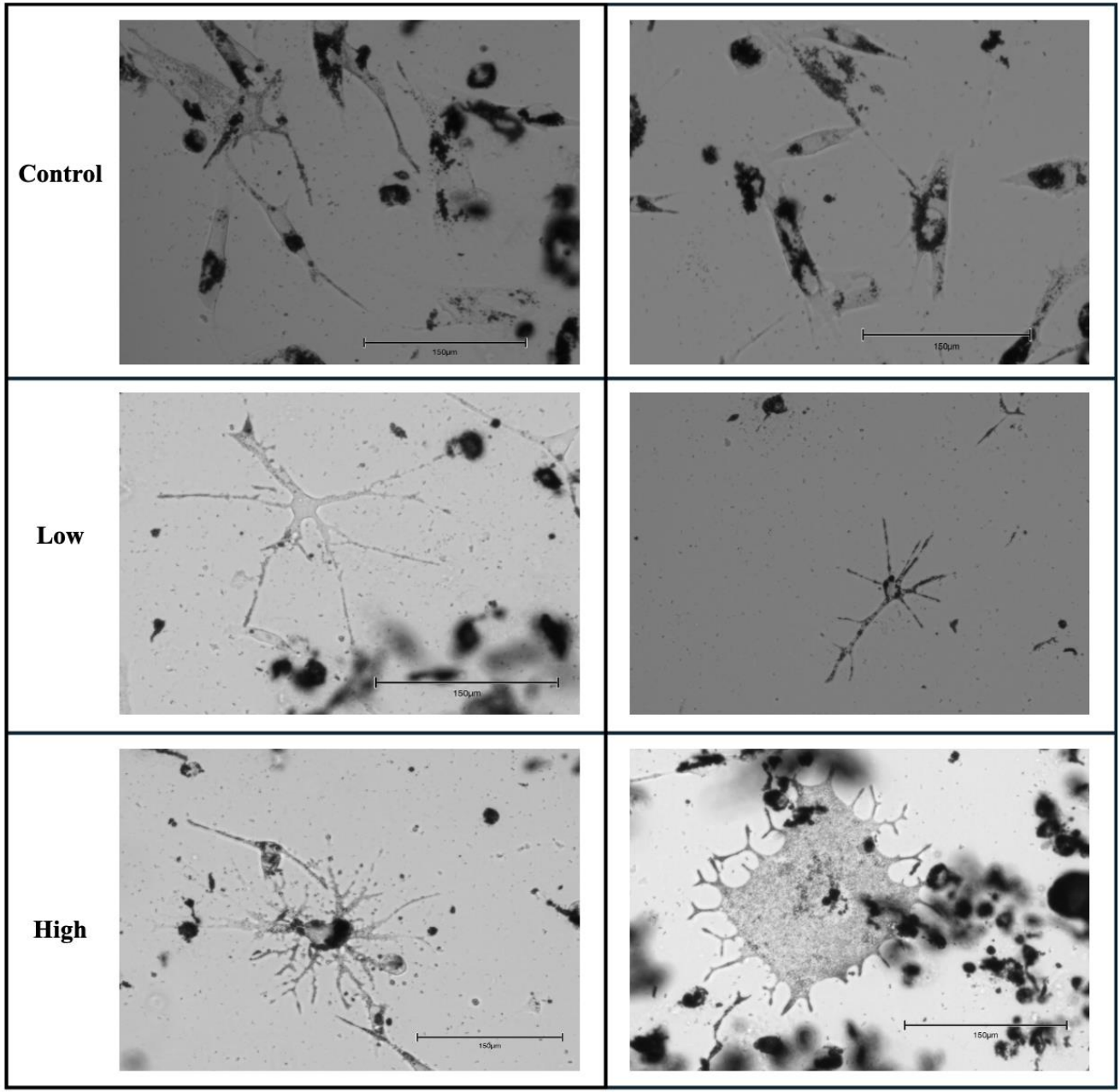

**Figure 4.** Photographs of representative B16-F10 mouse melanoma cells taken using an EVOS 5000 microscope in varying different conditions, control (0mM), low (1mM), and high (5mM) theophylline treatments with a total magnification of 200x. Cells exposed to a high concentration of theophylline showed multiple dendrites and different branching compared to the control and low conditions. The branching of the dendrites was increased proportionally to the concentration of theophylline the cells were exposed for each condition.

The control condition showed the highest number of total cells based on the MTT Assay results (Figure 5), and as illustrated on the photographs taken for the different conditions (Figure 4). Representative images of the different conditions showed distinct differences. The control condition has the lowest number of dendrites, if dendrites were present the majority were short. The addition of theophylline showed an increase in multiple dendrites and different branching compared to the control. Cell exposed to the highest concentration of theophylline (5mM) showed a greater number of dendrites and different branching compared to the control and low concentration (1mM) conditions (Figure 4). Thus, melanosomes were identified to be present at a greater concentration in dendrites of the cells exposed to the high concentration of theophylline than the low concentration treatment (Figure 4).

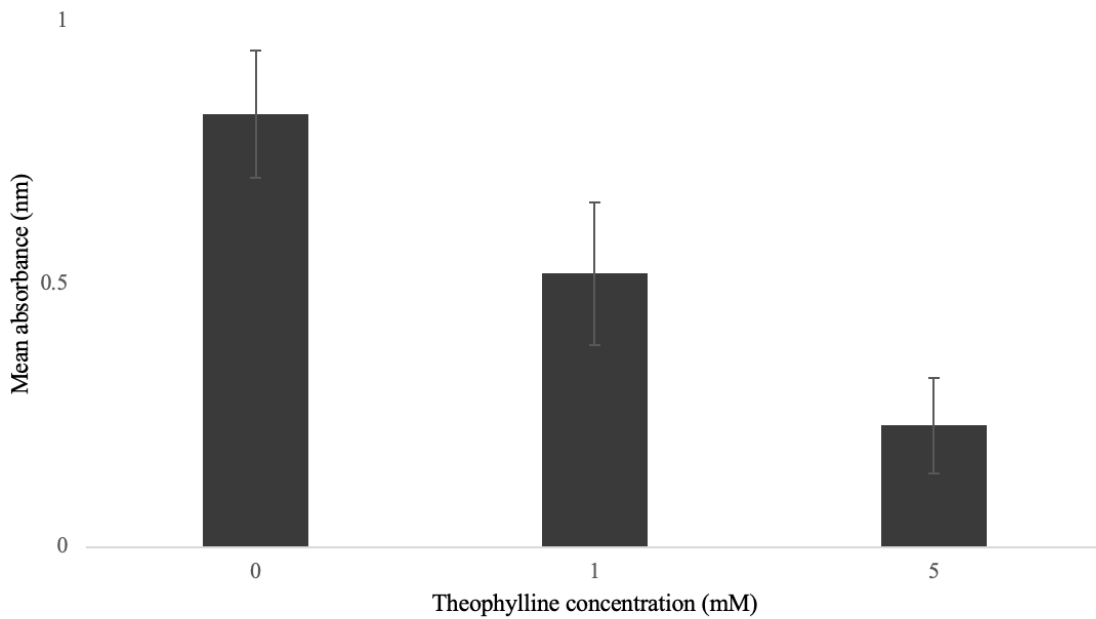

**Figure 5.** Mean absorbance values of the dissolved formazan in each well containing B16-F10 cells measured at 570 nm at temperature set point of 25˚C. Error bars show a ± standard error of the mean n=4; p = 0.544376.

**DISCUSSION**

The B16-F10 cells that were not treated for melanoma displayed traits of cells derived from undifferentiated melanocytes. They had lamellipodia, a noticeable Golgi apparatus, a considerable number of mitotic figures, and nucleoli that varied in number. Melanoma cells and other cancer cells are known for rapidly dividing, and it is an important factor that was consider for this study. The estimated doubling time of 0.67 hours was significantly less than what was expected. This estimate likely results from the incorporation of data points outside the true exponential growth phase due to experimental error in counting cells during the various collection data points. Previous research describes doubling times for B16-F10 cells of 20 to 30 hours under standard culture conditions (Tada et al., 2002). Therefore, our calculated rate likely overestimates actual proliferation and potentially requires reevaluation with more appropriate regression points or data exclusion criteria. As well as increasing sample size for to approach to the true value for doubling time.

Morphological results with theophylline-treated cells clearly show that elevated cAMP levels induced differentiation. Differentiation in this case refers to the establishment of specialized structures and functions in melanoma cells, specifically the development of more extensive and more numerous dendrites. The low and high concentration of theophylline conditions showed a unique morphology compared to control and untreaded B16-F10 mouse melanoma cells, which showed a lower number of dendrites in comparison to treated cells. Therefore, indicating that treated cells differentiated, showing proliferation by expressing longer and thinner dendrites containing a large number of melanosomes, which were not observed in the control BF16-F10 mouse melanoma cells. Theophylline concentration was correlated with dendritic complexity, as illustrated in Figure 4, with cells in the 5 mM treatment group displaying the most pronounced morphological alterations. When the levels of theophylline were increased cAMP levels were also increased. Thus, showing the correlation that an increase in cAMP levels decreases cell proliferation and increases cell differentiation. This aligns with the established function of cAMP in promoting melanocyte differentiation by inducing dendritic outgrowth and melanin synthesis (Alberts et al., 2023). The effects we observed confirm the hypothesis that intracellular stabilization of cAMP, via phosphodiesterase inhibition by methylxanthines like theophylline, increases cell differentiation on melanoma cells (Tada et al., 2002).

Parallel to the morphological alterations, the MTT assay also showed a dose-dependent decrease in cell proliferation, as reflected by lower values of absorbance in both 1 mM and 5 mM theophylline-treated groups compared to the control (Figure 5). According to a basic principle of cell biology, differentiated cells will exit the cell cycle and stop dividing, which is consistent with the inverse relationship between proliferation and differentiation (Alberts et al., 2023). However, based on the statistical analysis no significant difference was found between the control, low and high concentrations conditions ($p = 0.544376$), this could have been due to an earlier administered MTT assay test. While we inferred a cAMP increase from theophylline treatment, direct measurement was not carried out in our current experiment. For measuring cAMP levels, future studies could utilize enzyme immunoassays (EIA) or radioimmunoassay (RIA), which are widely used for sensitive quantification of intracellular cyclic AMP (Tada et al., 2002). FRET (fluorescence resonance energy transfer)-based biosensors also offer a real-time, non-destructive method for monitoring cAMP dynamics in intact, living cells.

Our findings and observed trends are consistent with previous research indicating that cAMP analogs and phosphodiesterase inhibitors can induce differentiation of melanoma cells and suppress their growth (Tada et al., 2002). This is consistent with the overall differentiation therapy approach, which aims to force cancer cells out of the proliferative cycle and into a mature, less malignant form. Initially demonstrated in leukemia models, differentiation therapy has become more popular as a potential strategy for treating solid tumors, including melanoma (Breitman, Selonick and Collins, 1980). Although not yet integrated into standard clinical practice for melanoma treatment, our data underscore the promise of further exploring cAMP-modulating agents as part of combination therapies to reduce tumor aggressiveness without relying solely on cytotoxic drugs.

Further investigations could measure cAMP directly and use different approaches to increase cAMP levels other than theophylline to clearly understand the role in cell proliferation of cAMP. Additionally, it would be intriguing to increase cAMP levels by using different methodologies to determine if changes in cAMP levels are an influence factor cell proliferation and cell differentiation.

**APPENDIX**

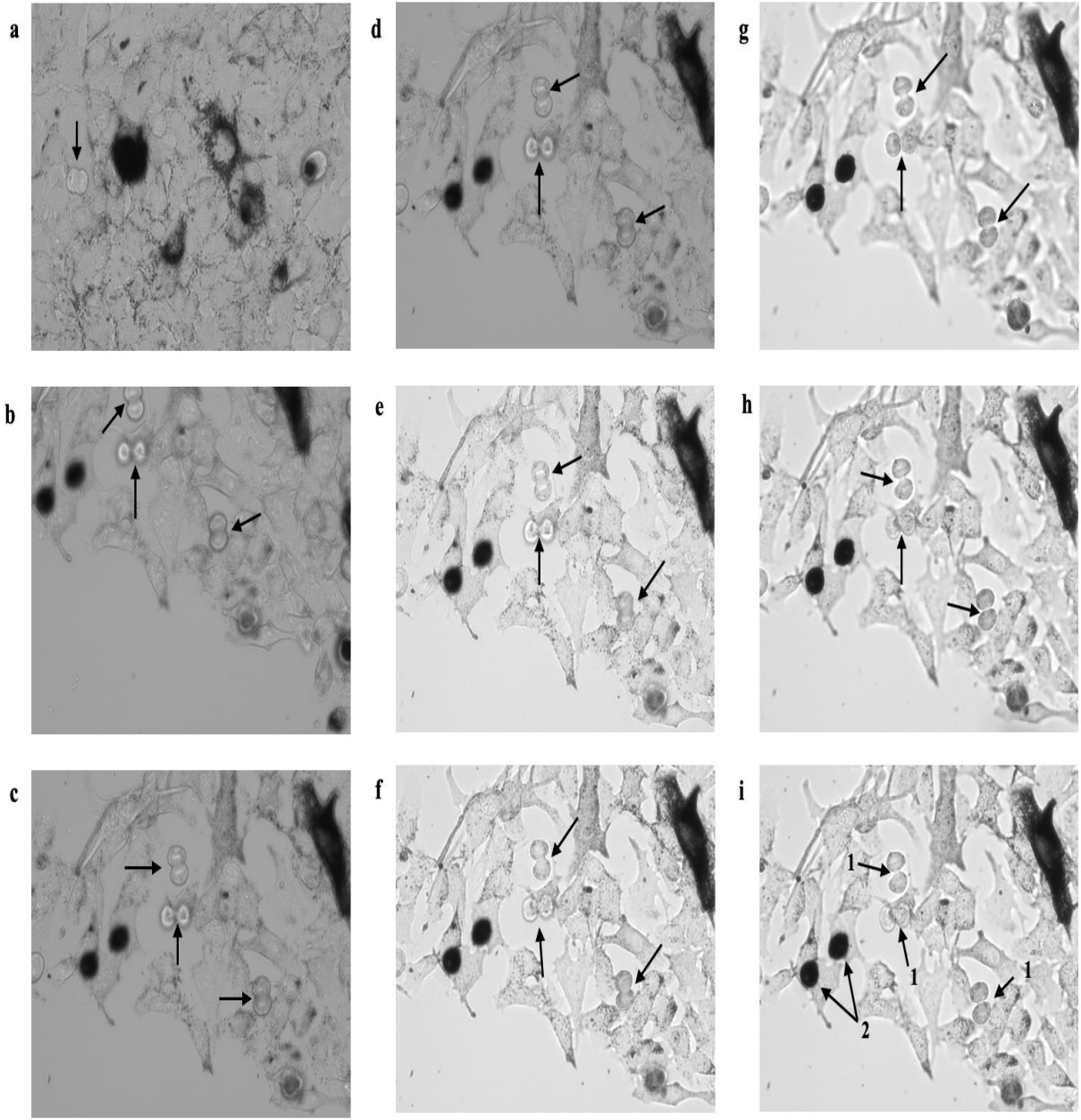

**Supplementary Figure 1.** Time-lapse of miotic cell division during transitioning stages to telophase and cytokinesis. Image a show a cell transitioning to early stages of telophase. Panels b-g depicts cells undergoing telophase in at different stages of telophase. Image h shows late stages of telophase. Image i shows cells undergoing completing cytokinesis (1), and cell two dully divided cells (2). Arrows emphasize the characteristics mentioned in each photograph.